\documentclass[journal=,manuscript=article,layout=twocolumn
]{achemso}

\pdfoutput=1
\usepackage{geometry}
\usepackage[normalem]{ulem}
\setkeys{acs}{articletitle=true}
\usepackage{lettrine}
\usepackage{comment}
\usepackage{xr}
\usepackage[version=3]{mhchem} 
\usepackage[T1]{fontenc}       

\usepackage[fontsize=10pt]{scrextend}
\usepackage{newtxtext}
\usepackage{bm} 
\usepackage[labelfont=bf]{caption} 
\usepackage[hyperfootnotes=false]{hyperref}	 
\hypersetup{
  colorlinks,
  citecolor=blue,
  linkcolor=blue,
  urlcolor=blue
}

\usepackage{titlesec} 
\titleformat{\subsection}[runin]
{\normalfont\bfseries}{\thesubsection{.}}{1em}{}[.]

\titleformat{\section}
{\normalfont\sffamily\large\bfseries}{\thesection{.}}{1em}{\MakeUppercase}

\usepackage[table]{xcolor}
\usepackage{graphicx}
\usepackage{amsmath,fullpage,amssymb,mathtools}
\usepackage{graphics,epsfig}
\usepackage{makeidx}
\usepackage{gensymb}
\usepackage{siunitx}
\usepackage{multirow}
\usepackage{appendix}

\setcitestyle{super}
\newcommand{\kT}{k_\text{B} T}
\newcommand{\ktot}{v}
\newcommand{\kapp}{k_{\text{app}}}
\newcommand{\koff}{k_{\rm off}}
\newcommand{\kr}{k_{\rm r}}
\newcommand{\kR}{k_{\rm R}}
\newcommand{\kd}{v_{\rm D}^0(\eps)}

\newcommand{\kon}{k_{\rm on}}

\newcommand{\eps}{\varepsilon_{\text{s}}}

\newcommand{\Nmax}{N_{\text{max}}}
\newcommand{\Rcat}{R_{\text{cat}}}

\newcommand{\rreac}{r_{\rm reac}}
\newcommand{\lb}{\lambda_{\text{B}}}
\newcommand{\fig}{Fig.~}
\newcommand{\Figs}{Figs.~}
\newcommand{\Eq}{Eq.~}
\newcommand{\Eqs}{Eqs.~}

\usepackage{soul}	

\SectionNumbersOn
\let\oldmaketitle\maketitle
\let\maketitle\relax

\title{Electrostatic reaction inhibition in nanoparticle catalysis}

\author{Yi-Chen Lin}
 \affiliation{Applied Theoretical Physics-Computational Physics, Physikalisches Institut, Albert-Ludwigs-Universit{\"a}t Freiburg, Hermann-Herder Strasse 3, D-79104 Freiburg, Germany}
\author{Rafael Roa}
\affiliation{Departamento de F{\'i}sica Aplicada I, Facultad de Ciencias, Universidad de M{\'a}laga, Campus de Teatinos s/n, E-29071 M{\'a}laga, Spain}
\author{Joachim Dzubiella}%
 \email{joachim.dzubiella@physik.uni-freiburg.de}
 \affiliation{Applied Theoretical Physics-Computational Physics, Physikalisches Institut, Albert-Ludwigs-Universit{\"a}t Freiburg, Hermann-Herder Strasse 3, D-79104 Freiburg, Germany}%
\alsoaffiliation{
 Research Group for Simulations of Energy Materials, Helmholtz-Zentrum Berlin, D-14109 Berlin, Germany
}%

\begin{document}
\pagenumbering{arabic}
\noindent

\parindent=0cm
\setlength\arraycolsep{2pt}

\twocolumn[	
\begin{@twocolumnfalse}
\oldmaketitle
\begin{abstract}
 Electrostatic reaction inhibition in heterogeneous catalysis emerges if charged reactants and products are adsorbed on the catalyst and thus repel the approaching reactants. In this work, we study the effects of electrostatic inhibition on the reaction rate of unimolecular reactions catalyzed on the surface of a spherical model nanoparticle by using particle-based reaction-diffusion simulations. Moreover, we derive closed rate equations based on approximate Debye-Smoluchowski rate theory, valid for diffusion-controlled reactions, and a modified Langmuir adsorption isotherm, relevant for reaction-controlled reactions, to account for electrostatic inhibition in the Debye-H{\"u}ckel limit.  We study the kinetics of reactions ranging from low to high adsorptions on the nanoparticle surface and from the surface- to diffusion-controlled limits for charge valencies 1 and 2. In the diffusion-controlled limit, electrostatic inhibition drastically slows down the reactions for strong adsorption and low ionic concentration, which is well described by our theory. In particular, the rate decreases with adsorption affinity, because in this case the inhibiting products are generated at high rate. In the (slow) reaction-controlled limit, the effect of electrostatic inhibition is much weaker, as semi-quantitatively reproduced by our electrostatic-modified Langmuir theory. We finally propose and verify a simple interpolation formula that describes electrostatic inhibition for all reaction speeds ('diffusion-influenced' reactions) in general. \end{abstract}
\end{@twocolumnfalse}]

\maketitle
\setlength\arraycolsep{2pt}
\small

\section*{Introduction}

Metallic nanoparticles (NPs) have received much attention in the past decade because of their extraordinary catalytic performance and other intrinsic properties at the nanoscale, \cite{haruta1997, bell:science, burda2005, astruc} and provide a wide range of potential applications, e.g., in biosensors, \cite{anker2008} fuel cells \cite{strasser2010} and other catalytic systems. \cite{crooks2001, lu2006, herves2012, wunder2010}
A critical issue of such a catalytic system is reaction inhibition, e.g., steric product inhibition if the products have a non-negligible adsorption affinity to the catalyst and block catalyst active sites,~\cite{Bowdenbook, gu2020, roa2018} which can drastically alter the kinetics and deactivate the catalytic performance. \cite{bhugun1996, horvath2004, luo2010, lang2014, hu2019} However, apart from steric interactions, reacting molecules and catalyzed products typically carry an electrostatic charge and thus involve long-ranged coupling between all species. Examples include prominent model reactions catalyzed by NPs, such as the reduction of (monovalent) nitrophenol or (trivalent) hexacyanoferrate. \cite{you2006, carregal-romero2010, hu2019} 

It was in fact shown in experiments that the electrostatic repulsions induced by the adsorbed borohydride (BH\textsubscript{4}\textsuperscript{-}) between the catalytic surface and reactants could slow down the reduction rate of 4-nitrophenol. \cite{choi2016} Hence, ionic reactant/product adsorptions and subsequent electrostatic interactions between the adsorbed species and the approaching reactants is an essential mechanism or reaction inhibition. The important role of electrostatic surface properties in nanoparticle catalysis was recently stressed by Roy {\it et al}.~\cite{roy} Here, also the question emerges how electrostatic rate regulation can be controlled by the presence of salt which screens electrostatic interactions. 

Consider, for example, an irreversible unimolecular reaction of an reactant A, transformed to B on a nanoparticle catalyst. The reaction mechanism can be written as 
\begin{equation}
\begin{split}
 \text{A} + \text{catalyst} &\xrightleftharpoons[\koff]{\kon} \text{A}_{\text{ad}} \xrightarrow{\kr} \text{B}_{\text{ad}}\\
 \text{B}_{\text{ad}} &\xrightleftharpoons[\kon ']{\koff '}  \text{B} + \text{catalyst},
\end{split}
 \label{eq:adsorption_chem_eq}
\end{equation}
where $\kon$ and $\kon'$ are the adsorption rate constants, $\koff$ and $\koff'$ are the desorption rate constants of reactant A and product B, respectively. The rate $\kr$ is the intrinsic (surface) rate constant that adsorbed reactant A converts to product B. Intermediate steps involve adsorbed reactants and products, $\text{A}_{\rm ad}$ and $\text{B}_{\rm ad}$. If reactant and products are charged, the adsorbed molecules will repel the approaching reactants and will decrease the rate. We call this effect {\it electrostatic reaction inhibition}, which is subject of this study. Quantitative details of such a fundamental effect depend on the nature of the reaction (e.g., diffusion- versus reaction-controlled reaction) and are not yet quantitatively understood. 

For example, a reaction-controlled (also 'surface-controlled') reaction rate $\ktot_{\rm R}$, valid for very slow chemical reactions, is typically simply proportional to surface adsorption, via \cite{Atkins} 
\begin{equation}
 \ktot_{\rm R} = \kapp \theta_{\text{A}}, 
 \label{eq:Langmuir_rate}
\end{equation}
where $\theta_{\text{A}}$ is the reactant surface coverage and $\kapp$ (units of ns\textsuperscript{-1}) is the apparent rate constant.
Conventionally, the Langmuir isotherm \cite{girgis1991}  (leading to a single-molecule Langmuir-Hinshelwood rate equation~\cite{ye2019}) provides a satisfactory description of surface coverage in the reaction-controlled limit \cite{bhugun1996, foo2010, yigit2012, hartvig2011} 
\begin{equation}
 \theta_{\text{A}} = \frac{K_{\text{eq}}\rho_{\text{A}}^0}{1 + K_{\text{eq}}\rho_{\text{A}}^0},
 \label{eq:Langmuir}
\end{equation}
where $K_{\text{eq}} = k_{\text{on}} / k_{\text{off}}$ is the equilibrium binding constant, and $\rho_{\rm A}^0$ is the bulk concentration of free A particles.  
\Eq\ref{eq:Langmuir} describes the adsorption of gases on surfaces with a finite number of binding sites in equilibrium. The equilibrium assumption is justified because reaction-controlled reactions are in a steady-state and distributions are very close to equilibrium. The basic Langmuir form can be easily extended to include the steric inhibition by adsorbed products by using multi-component Langmuir isotherms.\cite{laidler, Gomez2015}  Extensions to electrostatics have been also included in classical 'Stern-Langmuir' approaches for planar interfaces, as, e.g., in the binding of charged ligands to planar membranes~\cite{stern1924, davies1963, mclaughlin1981, dimov2002} or charged protein globules to microgels.~\cite{yigit2012}. However, in the realm of NP catalysis involving charged reactants/products adsorbed on spheres, such a correction and its consequences on the reaction rate has not been studied before. 

In the other limit, where chemical reactions are very fast and the total reaction only limited by diffusion, we have diffusion-controlled or diffusion-influenced nanoparticle-catalyzed reactions.\cite{angioletti2015, roa2017} Historically, Smoluchowski \cite{smoluchowski} established the theory for diffusion-controlled reaction rates. The Smoluchowski reaction rate constant for the diffusion-controlled limit in non-interacting systems, $k_0=4\pi D R$, depends only on the size of the nanoparticle, $R$, and the diffusion of reactants, $D$.  An  important extension was devised by Collins and Kimball to consider diffusion-{\it influenced} reactions, \cite{collins1949} where the surface reaction becomes significant, while other extensions followed in time.\cite{calef1983,berg1985,haenggi1990, joe2005, piazza2013} The important extension of Debye was to include (electrostatic) interactions between reactants and the associating partner.\cite{debye}
This Debye-Smoluchowski framework was employed, e.g., for electrostatics-driven macromolecular association kinetics on the Debye-H{\" u}ckel level of electrostatic treatments. \cite{sun2007, zhou_mccammon2008, yap2013, berezhkovskii2016}.  To the best of our knowledge, however, the Debye-Smoluchowski approach has not been extended yet to include self-consistently interactions between adsorbed products and approaching reactants. Self-consistency is needed because the steady-state distributions, in particular the product coverage, depends on the reaction rate itself. 

In order to check the typically approximative theories, particle-based reactive computer simulations have become a powerful tool to study complex reaction-diffusion systems.~\cite{ridgway2008, schoeneberg2014,froehner2018, manuel2019, weilandt2019, YiChenLin2020} Dibak {\it et al.} demonstrated, for example, that the results of particle-based simulations are consistent with the Debye-Smoluchowski theory.~\cite{manuel2019} The virtue of these simulations is that particle-particle interactions are well-defined and tunable. Electrostatic interactions can be included explicitly by Coulomb potentials, or screened Yukawa interactions in the presence of (implicit) salt. Moreover, the intrinsic and microscopic reaction rates are well-defined and can be clearly translated to theory~\cite{angioletti2015, froehner2018}. 

To summarize the motivation, despite the importance of nanoparticle catalysis for modern chemistry and the long history of theories and numerical approaches to fundamental reaction kinetics, the role of electrostatics in reaction inhibition is neither well understood nor put into quantitative descriptions.  In this article, we present closed approximate equations for electrostatic reaction inhibition for both diffusion-controlled and reaction-controlled reactions based on Debye-Smoluchowski and Langmuir approaches, respectively, and the Debye-H{\"u}ckel theory. The equations are verified by particle-based reaction simulations for investigating the feedback of adsorbed reactants and products in unimolecular reactions. We study the effects of the reactant concentration, the ionic strength and the adsorption affinity of the reactants/products to a spherical catalyst on the kinetics. Hence, this work provides a microscopic view of understanding the self-inhibition of catalyst kinetics due to electrostatic interactions. 

\section*{Simulation models and methods}

\begin{figure*}[htb!]
 \includegraphics[width = 0.65\textwidth]{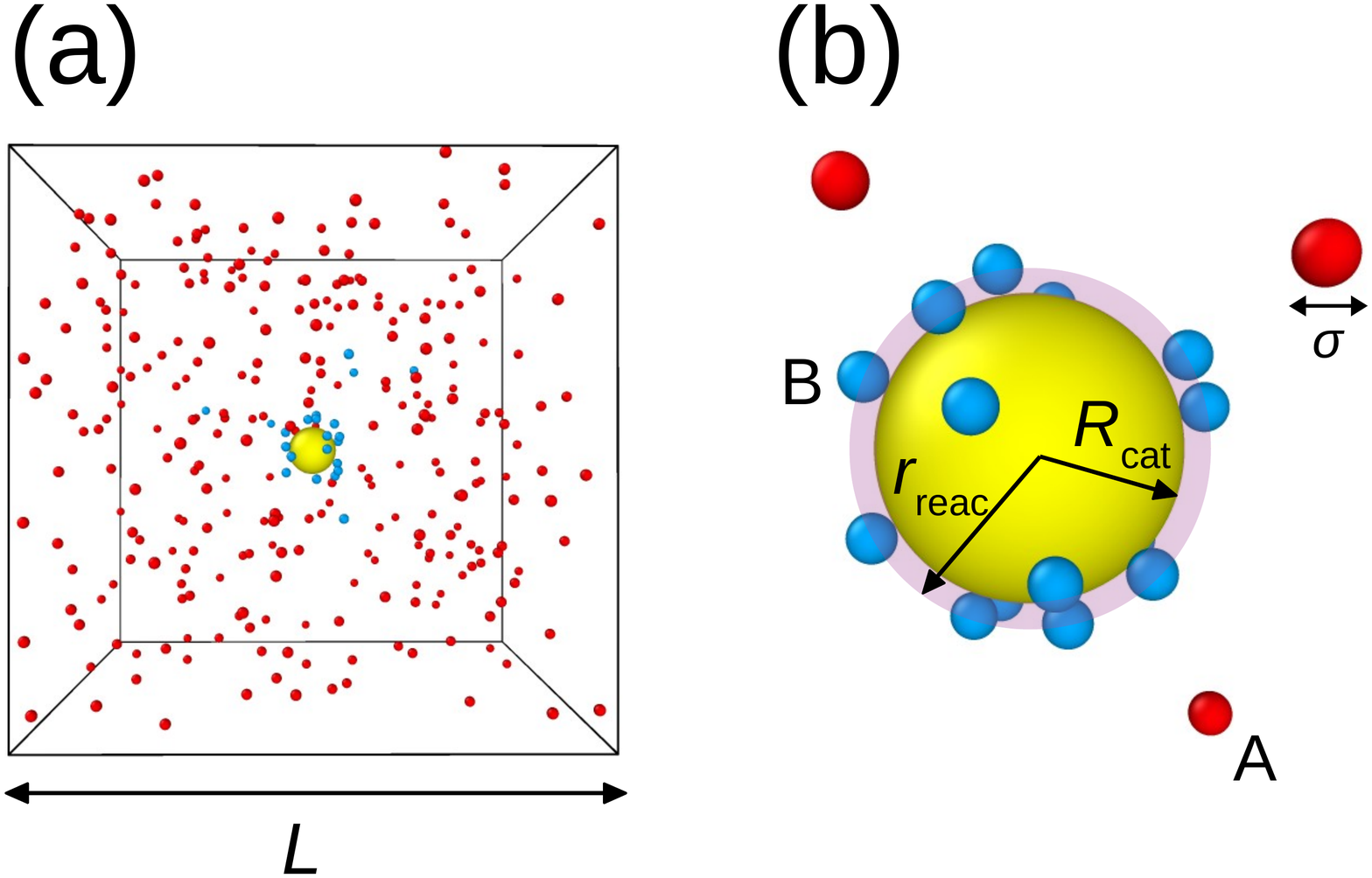}
 \caption{Simulation model and snapshots. (a) A nanoparticle catalyst (yellow sphere) is placed at the center of a cubic box of length $L$ with periodic boundary condition applied in all directions. Reactants (red particles) are randomly distributed in the cubic box at the beginning of the simulation. Reactants are transformed into products (blue particles) with reaction propensity $\kr$ when their distance from the nanoparticle catalyst of radius $\Rcat$  is smaller than the reaction cutoff radius $\rreac=\Rcat+2^{1/6}\sigma$, see panel (b). Reactants A (red) and products B (blue) have identical (Lennard-Jones) diameter, $\sigma$, and also identical (Lennard-Jones) adsorption energy to the nanoparticle, $\eps$.  To maintain steady-state, products are transformed back into reactants when they reach the boundary of the simulation box.}
 \label{fig:snapshot}
\end{figure*}

\subsection*{Model}

Our simulation model is depicted in \fig\ref{fig:snapshot}. We consider a solid perfect sphere representing a nanoparticle catalyst of radius $\Rcat =1$~nm at the center of a cubic box of length $L=20$~nm. We include in our simulation box an initial number of reactants, $N_{\text{A}}^0=300$. Therefore, the initial average density of reactants $\rho_{\rm A}^\text{initial}$ is 0.0375~nm$^{-3}$.  All reactants, A, and products, B, are mutually interacting with the Lennard-Jones (LJ) pair potential 
\begin{equation}
U_{\textrm{LJ}}(r_{ij}) = 4\varepsilon_{ij}\left[\left(\frac{\sigma}{r_{ij}}\right)^{12} - \left(\frac{\sigma}{r_{ij}}\right)^{6}\right],
\end{equation}
and a screened Coulomb potential (Yukawa)\cite{hansen2013}
\begin{equation}
\beta U_{\text{Yukawa}}(r_{ij}) = Z_{i}Z_{j}\lambda_{\text{B}}\frac{\text{exp}\left[-\kappa(r_{ij} - \sigma)\right]}{r_{ij}(1 + \kappa\sigma / 2)^2}, 
\label{eq:yukawa}
\end{equation}
between particles $i$ and $j$ in pair distance $r_{ij}$. We take the same diameter, $\sigma=0.4$~nm, for reactants and products. Moreover,
 $\varepsilon_{ij}=0.1~\kT$ is the LJ interaction energy, $\beta = 1 / \kT$, and $\lambda_\text{B} = 0.71$~nm is the Bjerrum length at temperature $T=298$~K. We consider that all particles have the same valency, $Z_{\rm A} = Z_{\rm B} = Z$ and we consider the two cases $Z=1$ or 2. We also include an implicit salt by the screening factor in the interaction given by Eq. \ref{eq:yukawa}, without explicit counterions, and chose the inverse Debye screening length, $\kappa=\sqrt{8\pi\ \lb N_{\rm Av}\times 10^3 I}$, where $N_{\rm Av} = 6.02 \times 10^{23}$ mol\textsuperscript{-1} is the Avogadro number, such that the ionic strength, $I$, ranges between 0.025 and 0.6~M. This corresponds to screening lengths $\kappa^{-1} = 0.39 \thicksim 1.93$ nm. The interaction between the catalyst and the particles is considered by a shifted LJ potential $U_{\text{LJ}}^{\text{s}}(r_i) =U_{\textrm{LJ}}(r_i - \Rcat)$, where $r_i$ is the distance of particle $i$ to the catalyst center, with an interaction energy $\varepsilon_\text{s}$ equal for reactants and products. The energy $\varepsilon_\text{s}$ essentially acts as an inverse temperature for the adsorption/desorption equilibrium and we change it between 0.1 and 10~$\kT$ to tune the adsorption.

The reactants A diffuse freely in the cubic box with diffusion constant $D$ and can reversibly adsorb (and desorb) on the catalyst surface subject to the shifted LJ potential $U_{\text{LJ}}^{\text{s}}(r_i)$. We define that adsorption (desorption) occurs if the particle distance $r_i$ is smaller (larger) than the reaction cutoff $\rreac=\Rcat+2^{1/6}\sigma$. In the adsorbed state, a reactant A can react according to the scheme given by \Eq\ref{eq:adsorption_chem_eq}. The resulting product B has also a diffusion constant $D$ and is subjected to the same shifted LJ potential. 
To reach the steady-state, products are transformed into reactants when they reach the edges of the simulation box.

\subsection*{Simulation basics}

We perform Brownian Dynamics (BD) simulations including molecular reactions. The particles A and B are explicitly resolved as diffusing solutes in a viscous continuum background (the solvent). The position of the $j$th particle $\bm{X}_j$ at time $t$ is computed by numerically iterating the overdamped Langevin equation using an Euler--Maruyama scheme \cite{ermak1978}
\begin{equation}
\gamma\frac{\partial \bm{X}_j(t)}{\partial t} = - \nabla_j U_{\rm tot} + \bm{R}_j(t),
\end{equation}
where $\gamma$ is the friction constant, $U_{\rm tot} = \sum_i [U_{\rm LJ}(\bm{X}_i) + U_{\rm LJ}^{\rm s}(\bm{X}_i) + U_{\rm Yukawa}(\bm{X}_i)]$ is the total interaction potential, and $\bm{R}_j(t) = (R_j^x, R_j^y, R_j^z)$ is a Gaussian random force, which satisfies the fluctuation--dissipation relation $\langle R_j^\alpha(t) R_j^\beta(t')\rangle = 2\gamma\kT\delta_{\alpha\beta}\delta(t-t')$ and has zero mean $\langle \bm{R}^\alpha_j(t)\rangle_t = 0$. In the simulations, we set units for the energy to $\kT$, the unit length to the nanometer, and the unit time to the Brownian time scale $\tau_{\rm B}$, such that the diffusion constant is $D = \kT / \gamma = 1$~nm$^2/\tau_{\rm B}$. The simulation time step is $\Delta t = 5\times 10^{-5}~\tau_{\rm B}$, and the simulations are performed up to 5000 $\tau_{\rm B}$. As a cutoff length of the LJ and the shifted LJ potential, we use $2.5\sigma=1$~nm and $\Rcat + 2.5\sigma = 2$~nm respectively.

\subsection*{Reactions in simulations}
When the distance of a reactant from the center of the catalyst is smaller than the reaction cutoff $\rreac$, an irreversible reaction, $\text{A}_\text{ad}\rightarrow\text{B}_\text{ad}$, occurs with the Poisson probability,  $P = 1 - \text{exp}(-\kr\Delta t)$, of finding at least one reaction event with rate (reaction propensity) $\kr$ in the time window $\Delta t$.~\cite{schoeneberger2013} We use $\kr$ as a free parameter to interpolate between diffusion-controlled and reaction-controlled reactions. For surface reactions, $\kr$ will play the role of a true surface rate constant (related to dimensionless surface coverage), as presented in the next theory section.  A similar numerical setup was used also in our previous study on bimolecular reactions catalyzed on the surface of a catalytic sphere.\cite{YiChenLin2020}

For a given $\kr$, in order to judge whether we deal with fast, more diffusion-controlled, or slow, more reaction-controlled  reactions in our systems, we need to define a surface rate constant with respect to reactant bulk concentration: our reaction takes place in a thin, spherical shell volume which we set to be $V_{\rm s} = (4\pi/3) [r_{\rm reac}^3 - (R_{\rm cat}+\sigma)^3)]$, confined by the reaction cut-off $r_{\rm reac}$ and the excluded volume of the nanoparticle sphere. With that we can define the surface rate constant $\kR = \kr V_{\rm s}$ with respect to reactant bulk concentration $\rho^0_{\rm A}$, in the sense that $k_{\rm R}\rho^0_{\rm A}=\kr V_{\rm s} \rho^0_{\rm A}$ provides the number of reactants reacting per unit time in the nanoparticle reactive shell without interactions. \cite{angioletti2015} The rate constant $\kR$ is then defined consistently with the Collins-Kimball surface rate constant in diffusion-influenced reactions.~\cite{collins1949,shoup1982,berg1985} We compare this to the fastest limit, that is, the Debye-Smoluchowski diffusion rate constant $k_0$ without electrostatic interactions~\cite{calef1983,berg1985} 
\begin{equation}
 k_0(\eps) = \left[\int_{r_{\rm reac}}^{\infty}\frac{e^{\beta U^{\rm s}_{\rm LJ}(r)}}{4\pi Dr^2}\text{d}r\right]^{-1} = 4\pi D R(\eps),
 \label{eq:kd}
\end{equation}
defining the effective (Smoluchowski) reaction radius $R(\eps) = [\int_{\rreac}^{\infty}{(\exp[{\beta U_{\rm LJ}^{\rm s}(r)}]}/{r^2})\text{d}r]^{-1}$. Due to the attraction in $U^s_{\rm LJ}$ the effective radius is slightly larger than $r_{\rm reac}$ and increasing with $\eps$, as summarized in  Table~\ref{tab:R}.  Finally, the total rate (absolute number of reactions per time) is then given according to Collins and Kimball by ${\ktot}^{-1}  = {(k_0\rho^0_{\rm A})}^{-1} + {(\kR\rho^0_{\rm A})} ^{-1}$. For very small propensities (slow 'chemical' surface transformations), thus $\ktot \simeq \kR\rho^0_{\rm A}$, for very large propensities, $\ktot \simeq k_0\rho^0_{\rm A}$. 

\begin{table}
\begin{tabular}{|c|c|c|c|c|c|c|} 
\hline
$\beta\eps$ & 0.1 & 1 & 3 & 5 & 7 & 10 \\
\hline
$R$ [nm] & 1.45 & 1.55 & 1.66 & 1.72 & 1.77 & 1.81 \\
\hline
\end{tabular}
\caption{Effective reaction (Smoluchowski) radius $R$ according to definition equation \ref{eq:kd}.}
\label{tab:R}
\end{table}

In our work, we choose propensities $\kr =$ 1.75 ${\tau_{\rm B}}^{-1}$, 17.5 $~{\tau_{\rm B}}^{-1}$, $\infty$, such that we simulate the rate constants in units of $k_0(\eps = 0) = k_0'$ as  $\kR \simeq $ 0.12 $k_0'$, 1.2 $k_0'$, and $\infty$ which we categorize as nearly reaction-controlled, diffusion-influenced and diffusion-controlled, respectively: While strictly a reaction-controlled reaction should obey $\kR \ll k_0'$, we consider $\kR = 0.12~k_0'$ as the slowest intrinsic rate constant in our simulations in concern of sampling quality: very slow reactions are simply not possible to sample in the available computational time. To simulate a diffusion-controlled reaction, $\kR = \infty$, we impose a reaction probability $P=1$, which in practice (regarding our finite time step simulation) leads to $\kR = 1372~k_0'$. To consider diffusion-influenced reactions, where both processes are of importance, we consider $\kR = 1.2~k_0'$.

As we argued above, the rate $\kd:=k_0 \rho_\text{A}^0$,  with $\rho_\text{A}^0$ the bulk reactant concentration,  is the fastest possible absolute rate. Therefore, we use it as a reference to scale our results of the total rate. The (steady-state) bulk density $\rho_\text{A}^0$ is an output from our finite-size simulations in the canonical ensemble. We obtain it by fitting the reactant radial density distribution functions with the analytically known long-range limits (cf. \Figs S1--S4). Note that $\rho_\textrm{A}^0$ depends on the interaction energy $\varepsilon_\text{s}$ because we simulate a canonical system with a fixed number of particles. The diffusion rate $\kd$ is expected to be close to the simulated reaction rate $k_{\rm R} \rightarrow\infty$ and no electrostatic interactions ($I\rightarrow\infty$), i.e., for instantaneous surface reactions without electrostatic reaction inhibition. 

The total absolute reaction rate $\ktot$ (events per time) in the simulations is obtained from the slope of the cumulative number of reaction events between $t = 0.2\ t_{\text{max}}$ and $t = t_{\text{max}}$, where $t_{\text{max}}$ is the maximum simulation time. 

\section*{Reaction rate theory including electrostatic product inhibition}

\subsection*{Diffusion-controlled reactions}
In regard of a theoretical framework, we start considering the rate of a diffusion-controlled reaction (valid for $\kR\rightarrow\infty$) at a single reactive catalytic sphere of Smoluchowski radius $R$ interacting with the reactants, which is based on the Debye-Smoluchowski equation\cite{debye}
\begin{equation}
 \ktot_{\rm D} = {\rho_{\rm A}^0}\left[\int_{R}^{\infty}\frac{e^{\beta V_{\rm el}(r)}}{4\pi Dr^2}\text{d}r\right]^{-1},
 \label{eq:Debye}
\end{equation}
where $V_{\text{el}}(r)$ is the electrostatic interaction potential between the product-covered catalyst and the reactants, and $D$ is the reactant diffusion coefficient, which we assume to be constant and position-independent. In a diffusion-controlled reaction a reactant A is quickly transformed into a product B, which leads to accumulation of adsorbed products on the catalytic surface. This makes the product surface concentration larger than the reactant bulk concentration. \cite{roa2018} The important consequence is an electrostatic repulsion between adsorbed products and approaching reactants which limits the rate of reaction at the catalyst surface.

Our theory is based on a perturbation approach for weak electrostatic interactions and stationary situations. We start by assuming that the total surface charge $Q$ contributed by $n$ adsorbed products can be written as
\begin{equation}
Q = Q_{\text{B}}n = Q_{\text{B}}\frac{\ktot_{\rm D}}{k_{\text{off}}},
\end{equation}
where $Q_{\text{B}}=Z_\text{B}e$ is the charge of a product, and $e$ being the elementary charge. The adsorption number $n$ comes from the stationary reaction with adsorption rate $\ktot_{\rm D}$ (which in this limit is equal to the total reaction rate), and a desorption rate $\koff$, the latter also in units number per time. In the stationary case, the generation of products and their desorption are balanced and $n$ becomes stationary. We assume that the desorption (or dissociation) constant $\koff$ is independent from ionic strength $I$, because it should be dominantly governed by the activated escape from the potential well with depth $\eps$. We will examine the validity of that assumption {\it a posteriori}. 

For weak electrostatics, the electrostatic interaction $V_{\text{el}}(r)$ which a reactant experiences from the adsorbed products, see \fig \ref{fig:snapshot}(b), is on a Debye-H{\"u}ckel level given by
\begin{equation}
\beta V_{\text{el}}(r) = \beta Z_{\rm A} e \psi(r) = n\cdot Z_{\rm A}Z_{\rm B}\lb\frac{e^{-\kappa(r - R)}}{r(1 + \kappa R)}, 
\label{eq:el}
\end{equation}
where $\psi(r)$ is the electrostatic potential generated by the adsorbed products. Another assumption made implicitly by writing equation \ref{eq:el} is that the the Smoluchowski radius $R$ also serves as defining the effective DH radius where we read off the surface potential. This assumption is very reasonable as we judge from radial density profiles (see for example Fig.~S3 in the Supporting Information) which feature a minimum at $R$ beyond the first 'solvation' peak of the reactants/products, thus correctly demarcating the complete adsorbed charged layer.  

Assuming that  that the electrostatic interaction is weak (i.e., we can linearize the exponent), the adsorption rate $\ktot_{\rm D}$ reads as
\begin{align}
  {\ktot_{\rm D}}^{-1} & = \int_{R}^{\infty}\frac{e^{\beta V_{\text{el}}(r)}}{4\pi Dr^2}\text{d}r\cdot{\rho_{\rm A}^0}^{-1} \nonumber \\ 
                     & = \left[\int_{R}^{\infty}\frac{1}{4\pi Dr^2}\text{d}r + n\cdot E(R,\kappa)\right]{\rho_{\rm A}^0}^{-1} \nonumber \\
                     & = ({v_{\rm D}^0})^{-1} + \frac{\ktot_{\rm D}}{\koff\rho_{\rm A}^0} E(R,\kappa).
 \label{eq:kon_1}
\end{align}

The electrostatic perturbation $E(R,\kappa)$ is thus defined as 
\begin{align}
  E(R, \kappa) & = \int_{R}^{\infty}\frac{\beta V_{\text{el}}(r)}{4\pi Dr^2n}\text{d}r \nonumber \\ 
                            & = \frac{Z_{\rm A}Z_{\rm B}\lb e^{\kappa R}}{4\pi D (1 + \kappa R)}\int_{R}^{\infty}\frac{e^{-\kappa r}}{r^3}\text{d}r.
\end{align}

By rearranging  \Eq\ref{eq:kon_1}, one can obtain an explicit, self-consistent form of the adsorption rate 
\begin{equation}
\begin{split}
 \ktot_{\rm D} = &-\frac{\koff\rho_{\rm A}^0}{2v_{\rm D}^0E(R, \kappa)} \\
                 &+ \sqrt{\left(\frac{\koff\rho_{\rm A}^0}{2v^0_{\rm D}E(R, \kappa)}\right)^2 + \frac{\koff\rho_{\rm A}^0}{E(R, \kappa)}}.
 \label{eq:kon_2_appendix}
\end{split}
\end{equation}
Considering again weak electrostatics, for small $E(R, \kappa)$, \Eq\ref{eq:kon_2_appendix} can be approximated (see the derivation in the Supporting Information) by
\begin{equation}
 \ktot_{\rm D}  \approx \frac{v_{\rm D}^0}{1 + \frac{(v_{\rm D}^0)^2}{k_{\text{off}}\rho_{\rm A}^0}E(R, \kappa)},
\label{eq:k_fit}
\end{equation}
which clearly shows the inhibition action of the rate by electrostatics: larger product adsorption or larger valencies increase $E$ in the denominator and the total rate decreases. Note that for large ionic strength, $\kappa \rightarrow \infty$, the electrostatic perturbation $E(R , \kappa)\rightarrow 0$ and the adsorption rate (cf. \Eq\ref{eq:k_fit}) recovers the Smoluchowski rate, $\ktot_{\rm D} \rightarrow \kd = k_0\rho^0_{\rm A} = 4\pi D R \rho^0_{\rm A}$, as needed. In experiments, $\kd$ can be measured at high salt concentrations.  

We will use \Eq\ref{eq:k_fit} to compare to our simulation results. The only fitting parameter is $\koff$, the desorption rate of the products from the nanoparticle. As argued above, in the fitting we assume that $\koff$ is independent from ionic strength. 

\begin{figure*}[ht!]
\includegraphics[width = 0.7\textwidth]{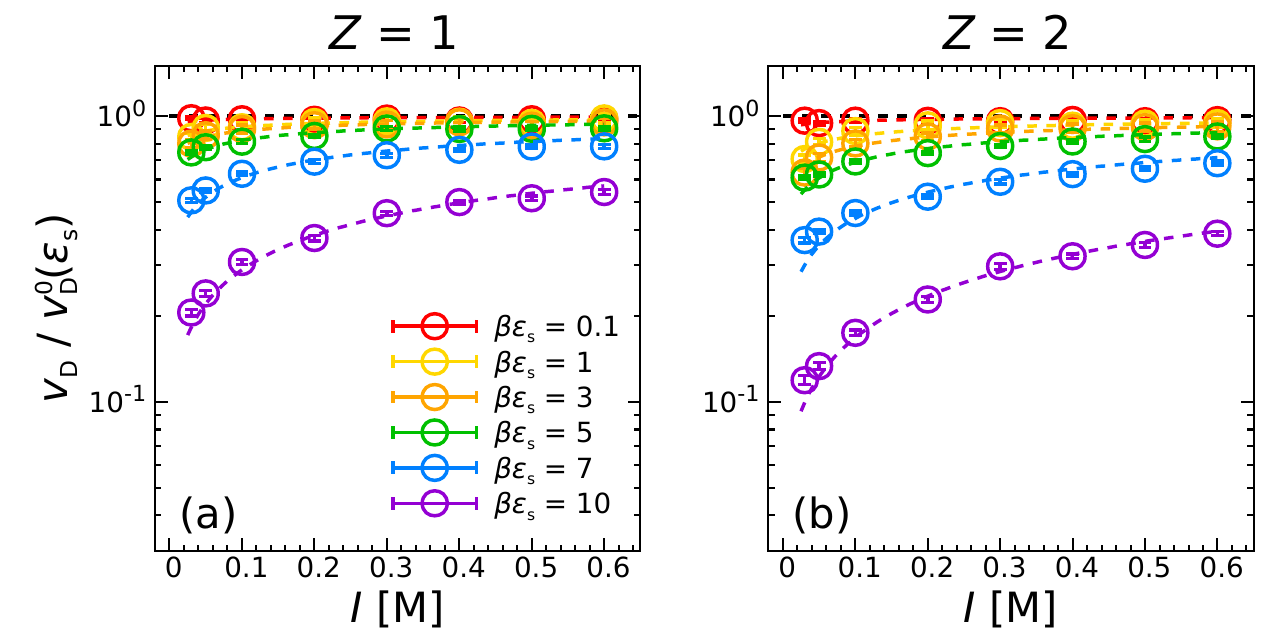}
\caption{Normalized total reaction rate $\ktot_{\rm D}$ in the diffusion-controlled limit ($k_{\rm R} = \infty$) as a function of the ionic strength $I$ 
for different reactant and product valencies (a) $ Z_{\rm A} = Z_{\rm B} = Z=1$, (b) $Z_{\rm A} = Z_{\rm B} = Z =2$. 
The open colored circles show simulation results for different adsorption energies $\varepsilon_\text{s}$. The dashed colored lines are the theoretical predictions given by \Eq~\ref{eq:k_fit}. The horizontal black dashed line represents the Smoluchowski-limit of the diffusion rate without electrostatic interactions, $v_{\rm D}^0(\eps)$.}
\label{fig:kapp_k_inf}
\end{figure*}
\subsection*{Reaction-controlled reactions}

Reaction-controlled reactions (valid for $\kR\rightarrow 0$) are slow and independent of the rate of diffusive processes, and we can devise a theory based on equilibrium distributions. We again start with the total reaction rate, $\ktot$, which in unimolecular surface reactions is typically defined as\cite{laidler, Atkins}
\begin{equation}
\ktot_{\rm R}
=\kr N_{\text{A}_\text{ad}} = \kr \theta_\text{A} \Nmax,
\label{eq:ktotsurf}
\end{equation}
where $N_{\text{A}_\text{ad}}$ is the number of adsorbed reactants. Typically surface reactions are phrased in terms of surface coverage which are described by Langmuir isotherms. Then, $\kr$ plays now the role of a true surface rate constant. Here we follow the same route to be close to the familiar literature. Then, $\Nmax$ in \Eq\ref{eq:ktotsurf} is the maximum adsorption number and $\theta_\text{A}$ is the reactant surface coverage defined as $\theta_\text{A} = N_{\text{A}_\text{ad}} / \Nmax$. 
In the Langmuir picture, $\Nmax$ describes the maximal number of potential binding sites. In our model, this limit is imposed by the maximal packing of the reactant particles on the nanoparticle surface. While this is strictly speaking different from the Langmuir model assumptions, it was demonstrated that the Langmuir model is well applicable and valid for not too high packing fractions of adsorbers.~\cite{yigit2012} $\Nmax$ is calculated from our simulations by fitting the Langmuir model to binding isotherms for the neutral (i.e., uncharged reactants) reference cases in full equilibrium, and we obtain $\Nmax=43$.  (See details in the Supporting Information; the result is as expected very near to the estimate from a close-packing adsorption for 2D hard disks on the catalyst.)

In stationary equilibrium, we have following rate balances:
\begin{subequations}
 \begin{align}
  \frac{\text{d}N_{\text{A}_{\rm ad}}}{\text{d}t} = & \kon \rho_{\rm A}^0\Nmax(1 - \theta_{\rm A} - \theta_{\rm B}) - \kr \Nmax\theta_{\rm A}- \koff \Nmax\theta_{\rm A} = 0,  \label{eq:chem_eq_rc_1a} \\
    \frac{\text{d}N_{\text{B}_{\rm ad}}}{\text{d}t} =& \kon '\rho_{\rm B}^0\Nmax(1 - \theta_{\rm A} - \theta_{\rm B}) + \kr \Nmax\theta_{\rm A} - \koff '\Nmax\theta_{\rm B} = 0, \label{eq:chem_eq_rc_1b}
 \end{align}
\end{subequations}
where again $\rho_{\rm A}^0$ and $\rho_{\rm B}^0$ are the bulk reactant and product concentrations, respectively.
Defining the equilibrium adsorption/desorption constants for reactants A and products B as
\begin{equation}
  K_{\rm A} = \frac{\kon}{\koff},\quad K_{\rm B} = \frac{\kon'}{\koff'},
\end{equation}
and considering that the surface reaction is the rate-determining step in a reaction-controlled reaction, i.e., $\koff \gg \kr$ and $\koff' \gg \kr$, \Eqs\ref{eq:chem_eq_rc_1a} and \ref{eq:chem_eq_rc_1b} turn into the known, multi-component Langmuir isotherms\cite{laidler, Gomez2015} for A-B coadsorption
\begin{subequations}
 \begin{align}
  \theta_{\rm A} &= \frac{K_{\rm A}\rho_{\rm A}^0}{1 + K_{\rm A}\rho_{\rm A}^0 + K_{\rm B}\rho_{\rm B}^0} \label{eq:surfcovA}\\
  \theta_{\rm B} &= \frac{K_{\rm B}\rho_{\rm B}^0}{1 + K_{\rm A}\rho_{\rm A}^0 + K_{\rm B}\rho_{\rm B}^0}. \label{eq:surfcovB}
 \end{align}
 \label{eq:surfcov}
\end{subequations}
The equilibrium constants in our model are equal for reactants and products since they have the same size and interaction parameters and can be expressed as
\begin{equation}
 K_{\rm A} = K_{\rm B} = K_{\rm eq} = V_0\cdot e^{-\beta U},
 \label{eq:eq_const}
\end{equation}
where $V_0$ is the effective binding volume per binding site in the Langmuir picture.~\cite{Gilson,general,xiao,zhou2009} The Langmuir binding volume $V_0$ is a function of $\beta\eps$ and we calculate it for each adsorption energy by fitting the binding of neutral reactants to the one-component Langmuir isotherm (see the Supporting Information). For $\beta\eps = 1$, $V_0$ is well approximated by the shell volume divided by the saturation binding number, i.e., $V_0 = V_{\rm s} / \Nmax = 0.0288$~nm$^{3}$, i.e., as could be expected from the average available volume per particle if all $\Nmax$ 'sites' are occupied. 
The $V_0$  can be easily translated to the standard volume l/mol to refer to the standard energy of binding in experiments.~\cite{Gilson,general,xiao}

Furthermore, $U$ is the surface (binding) energy experienced by the reactants
\begin{equation}
U = -\eps + V_{\rm el}(R),
\end{equation}
where we consider in addition to the surface Lennard-Jones attraction $\eps$, the contribution of the surface electrostatic interaction $V_{\rm el}(R)$ on the Debye-H{\"u}ckel level given by \Eq\ref{eq:el}, which is now a function of the total surface (charge) coverage
\begin{equation}
\beta V_{\text{el}}(R) = \frac{Z_{\rm A}\lb \Nmax (Z_{\rm A} \theta_{\rm A}+Z_{\rm B}\theta_{\rm B})}{R (1 + \kappa R)} 
\label{eq:elsurf}
\end{equation}

To summarize, the total reaction rate in the limit of reaction-controlled reactions with electrostatic product inhibition, $\ktot_{\rm R} = \kr \theta_\text{A} \Nmax$, can be calculated by numerically solving the coupled \Eqs\ref{eq:surfcov}--\ref{eq:elsurf}, where $\Nmax=43$ is globally used for all systems as introduced above and $V_0$ is an $\eps$--dependent input parameter. This theory is essentially fit-parameter free as we deduced/verified missing parameters from simulations of the neutral reference cases. Experiments could follow such a procedure by measuring first at very high ionic strengths.  
%
%

\begin{figure*}[ht!]
\includegraphics[width = 0.7\textwidth]{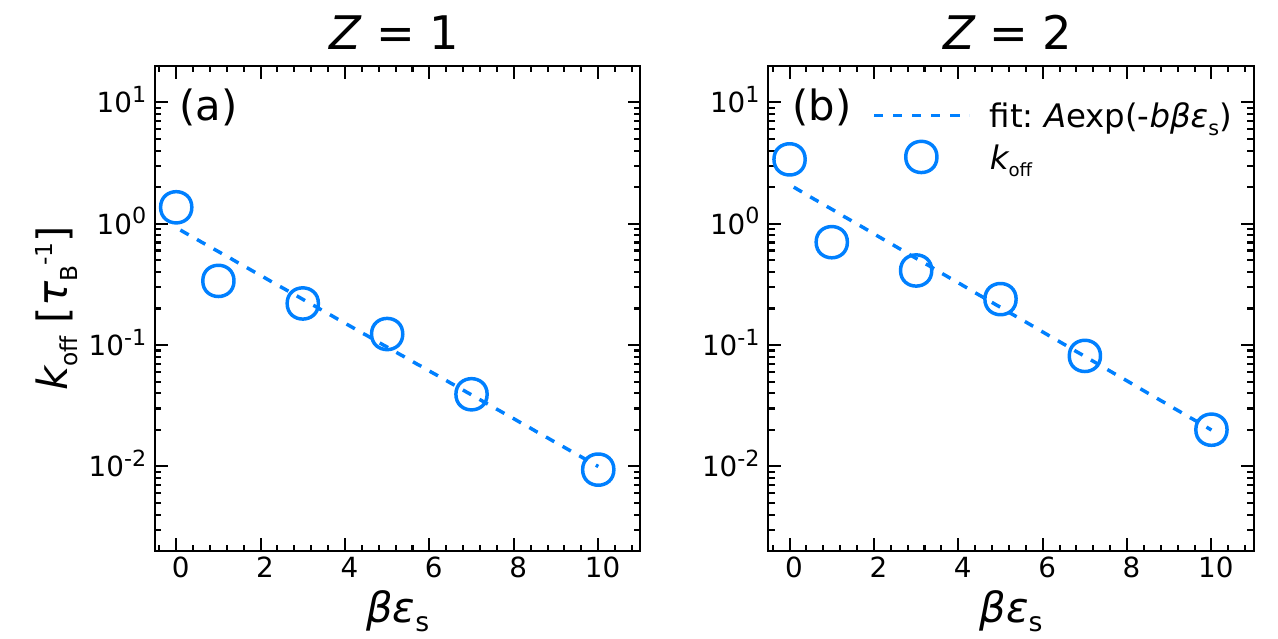}
\caption{Product desorption rate constant, $\koff$, as a function of the interaction energy $\eps$ in log-linear scale for (a) reactant/product valency $Z_{\rm A} = Z_{\rm B} = Z =1$ and (b) $Z_{\rm A} = Z_{\rm B} = Z =2$ for the diffusion-controlled reactions in \fig\ref{fig:kapp_k_inf}.
Open circles show the results of the fits of \Eq\ref{eq:k_fit} to the simulations. The dashed lines are exponential fits, $A\text{exp}(-b\beta\eps)$, of the obtained $\koff$ values with $A$ and $b$ as fitting parameters. We find $A = 0.9273$ and $b = 0.4532$ for $Z=1$ and $A = 1.9989$ and $b = 0.4607$ for $Z=2$.}
\label{fig:koff}
\end{figure*}

\subsection*{Diffusion-influenced reactions}

Both theories presented above cover either only the full diffusion-controlled limit ($\kR\rightarrow\infty$) or the full reaction-controlled limit ($\kR\rightarrow 0$) to describe electrostatic product inhibition. To model the situation in-between ('diffusion-influenced'), we simply assume the addition of the reciprocal diffusion-controlled and reaction-controlled rates
\begin{equation}
 \ktot^{-1} = {\ktot_{\rm D}}^{-1} + {\ktot_{\rm R}}^{-1}. 
 \label{eq:diffusion_influenced}
\end{equation}
in the spirit of the work by Collins \& Kimball for non-inhibiting systems. \cite{collins1949,shoup1982,berg1985}. Note that in the reference case of no electrostatics and no adsorption, where it should hold $V_0 = V_{\rm s} / \Nmax$, then  $\kr \theta_\text{A} \Nmax = \kr V_0 \rho_{\rm A}^0 \Nmax= \kr V_{\rm s} \rho_{\rm A}^0 = \kR\rho_{\rm A}^0$, and we indeed recover the correct Collins-Kimball limit for neutral, non-adsorbing situations.~\cite{berg1985, shoup1982} Including electrostatics, it is without a mathematical derivation not clear if relation \Eq\ref{eq:diffusion_influenced} holds, and we use it just for convenience for a simple interpolation. 

\section*{Results}

\subsection*{Diffusion-controlled reactions}

\fig\ref{fig:kapp_k_inf} shows the total reaction rate $\ktot_{\rm D}$ normalized by the diffusion rate without electrostatic interactions, $\kd$, as a function of the ionic strength, $I$. We analyze the results for different reactant/product interaction energies with the catalyst, $\varepsilon_\text{s}$, which we show by different colors. Panel~(a) considers a reactant/product valency $Z_{\rm A}=Z_{\rm B}=Z=1$ and panel~(b) shows $Z_{\rm A}=Z_{\rm B}=Z=2$. We display our simulation results by open circles. The theoretical curves shown by dashed colored lines are obtained by fitting the simulation results with \Eq\ref{eq:k_fit} with the desorption rate constant $k_\text{off}$ as the only fitting parameter. The black dashed horizontal line represents the fastest limit of the diffusion rate neglecting electrostatic interactions, $\kd$. 

Since reactants are immediately transformed to products upon adsorption, here we can talk about {\it electrostatic product inhibition}.
We see in  Fig.~\ref{fig:kapp_k_inf} that the latter in diffusion-controlled reactions is a substantial effect that slows down the reaction rate for large adsorption affinities and low ionic strengths. Since products and reactants have the same adsorption energy in our model, $\beta\eps$ determines the number of products that accumulate on the surface of the nanoparticle catalyst. For small interaction energy values, $\beta\eps\leq5$, the product surface coverage is very low (cf. \fig~S5 in the Supporting Information), which results in a very weak electrostatic repulsion between the reactants and the nanoparticle surface and a negligible rate inhibition. For large interaction energy values, $\beta\eps>5$, however, the catalyst surface is highly covered by products which not only hinders the surface by steric effects but also creates a strong electrostatic repulsion between the catalyst and the approaching reactants. (An enhanced reactant depletion compared to a neutral ideal gas is indeed visible in the steady-state radial density profiles, shown in \Figs~S1--S2 in the Supporting Information.)

The decrease of the reaction rate observed in \fig\ref{fig:kapp_k_inf} is more pronounced at low ionic strengths since the electrostatic repulsion between the reactants and the nanoparticle surface is less screened by the salt. The reaction rate is further slowed when the valency of reactants and products increases to $Z=2$, \fig\ref{fig:kapp_k_inf}(b), due to a stronger electrostatic repulsion. As an example, for very large attraction $\beta\eps=10$ and small ionic strengths with $Z=2$, the total reaction is almost one order of magnitude smaller than the diffusion rate without electrostatics. Even for very large ionic strengths, still a decrease by a factor 2 is found, pointing to significant remaining short-range repulsions. 

Moreover, the reaction rate is overall faster when the adsorption energy is small: Since faster reactions and larger adsorption energies result in significantly more accumulated products on the catalytic surface, we obtain stronger product inhibition being consistent with \Eq\ref{eq:kon_1} indicating a negative-feedback of high reaction rates. This leads to the interesting effect, that the total rate $\ktot_{\rm D}$ decreases with adsorption affinity for charged, strongly adsorbing products. (We sill see that this is in contrast with the reaction-controlled reaction where the rate always increases with adsorption.) Importantly, our approximative equation \Eq\ref{eq:k_fit} for the diffusion-controlled reaction describes the simulation data very well over the simulated parameter ranges. 
%
%

\begin{figure*}[h!]
\includegraphics[width = 0.7\textwidth]{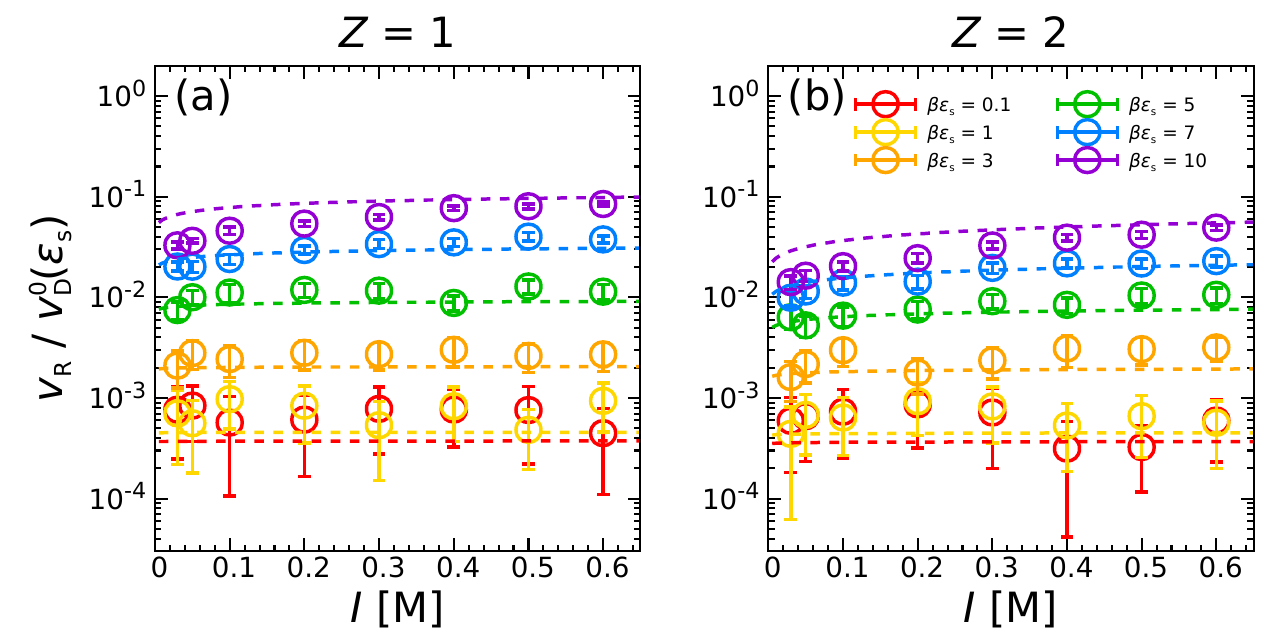}
%
\caption{Normalized total reaction rate $\ktot_{\rm R}$ for $\kR = 0.12~k_0'$ (assumed to be the slowest reaction constant in our simulations to probe the reaction-controlled limit) as a function of the ionic strength $I$ for different reactant and product valencies (a) $ Z_{\rm A} = Z_{\rm B} = Z=1$, (b) $Z_{\rm A} = Z_{\rm B} = Z =2$. The open colored circles show simulation results for different interaction energies $\varepsilon_\text{s}$. The dashed colored lines are the theoretical predictions given by \Eqs\ref{eq:surfcov}--\ref{eq:elsurf}.}
\label{fig:reaction_controlled}
\end{figure*}

The desorption rate constants $\koff$ obtained from the fitting procedure for different interaction energies $\eps$ and valencies $Z$ are displayed in log-linear plots by open circles in \fig\ref{fig:koff}. Panel~(a) in \fig\ref{fig:koff} shows the results obtained for $Z_{\rm A}=Z_{\rm B}=Z=1$ and panel~(b) for $Z_{\rm A}=Z_{\rm B}=Z=2$.  The dashed lines are exponential fits, $A\text{exp}(-b\beta\eps)$, of the obtained $\koff$ values with $A$ and $b$ as fitting parameters. We find that the desorption rate decreases indeed exponentially with increasing $\beta b\eps$ both for $Z=1$ and $Z=2$.  Since $\koff$ has dimensions of inverse of time, we interpret this quantity as the \textit{escape rates} of the adsorbed particles which, according to Kramers' escape rate theory, are proportional to the Boltzmann factor $\text{exp}(-\beta b\eps)$. \cite{haenggi1990} 
However, we find that, irrespectively of the interaction energy, the desorption rate $\koff$ for reactant/product valency $Z=2$ is two times larger than for $Z=1$. This valency dependence of the prefactor in Kramers' theory is not so easy to interpret and could be due to changes in mobility/noise or entropic effects.~\cite{Mila} Interesting is also the observation that the escape (or activation) energy is significantly smaller than $\eps$ by a scaling factor $b\simeq 0.45$ for both valencies. We suspect that the reason is electrostatic cooperativity between the products for larger adsorptions: a product particle feels the repulsion from other adsorbed products on the curved spherical particle which decreases the activation energy for escape. (The curvature leads to a net force radially out from the sphere.) This is supported by the data points for $\eps=1$ in \fig~\ref{fig:koff} in both panels (a) and (b), which both are significantly lower than the fitting line. For this very small adsorption case, the electrostatic contribution is essentially vanishing and thus the slopes in this plot (being the activation energy) closer to $\eps$. For intermediate to larger adsorptions, however, we observe a stretched exponential Kramers escape $\propto \exp[(-\beta\eps)^b]$, presumably due to electrostatic cooperativity.  If that is the case, in contrast to our assumption $\koff$ should also be ionic strength dependent, as in general the product adsorption numbers are (see \fig~S5 in the Supporting Information). However, it seems this dependence is not that strong, probably due to the self-regulation in the system. Hence, an average $\koff$-value as a single fit parameter for a system is sufficient to describe $\ktot_{\rm D}(I)$ and serve a useful purpose in fitting and extrapolating experimental data. 


%

\begin{figure*}[!h]
\includegraphics[width = 0.7\textwidth]{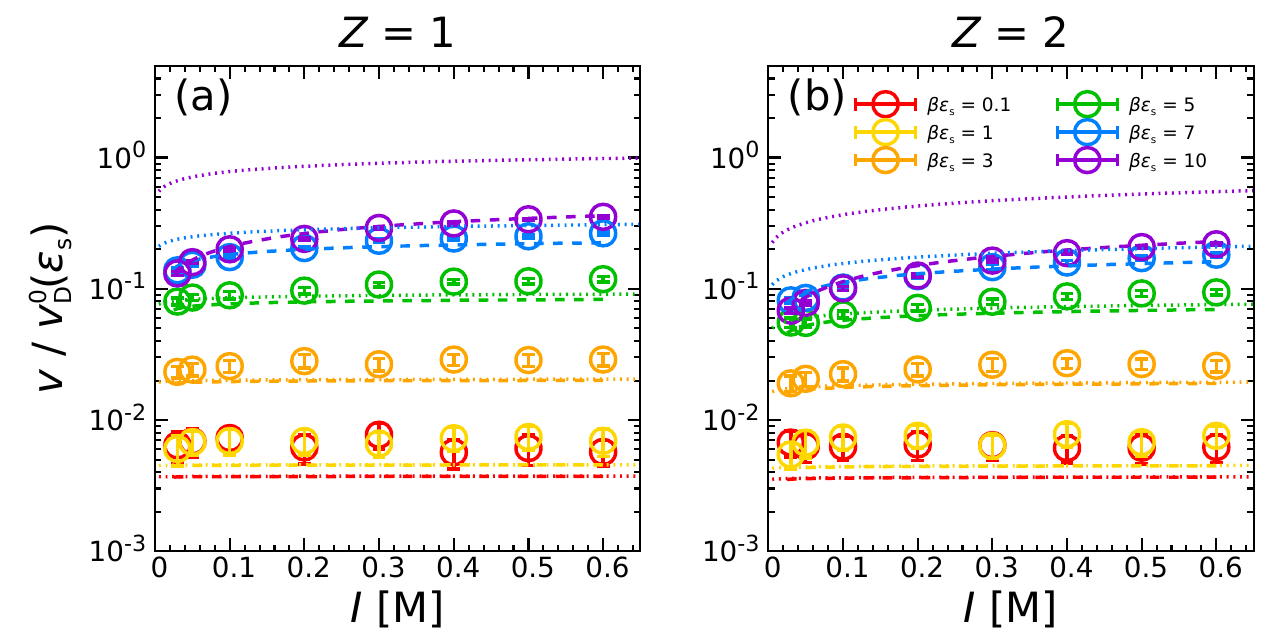}
\caption{Normalized total reaction rate $\ktot$ for $\kR = 1.2~k_0'$ as a function of the ionic strength $I$ for different reactant and product valencies (a) $ Z_{\rm A} = Z_{\rm B} = Z=1$, (b) $Z_{\rm A} = Z_{\rm B} = Z =2$. The open colored circles show simulation results for different interaction energies $\varepsilon_\text{s}$. The dotted and dashed colored lines are the theoretical predictions of the reaction-controlled theory, \Eqs\ref{eq:surfcov}--\ref{eq:elsurf}, and the diffusion-influenced theory, \Eq\ref{eq:diffusion_influenced}, respectively.}
 \label{fig:kapp_k_10kD}
\end{figure*}

\subsection*{Reaction-controlled limit}

\fig\ref{fig:reaction_controlled} shows the reaction-controlled limit (dashed lines), which is calculated by \Eqs\ref{eq:surfcov}--\ref{eq:elsurf}, and the simulation data (open circles) of the total rate $\ktot_{\rm R}$ with $\kR = 0.12~k_0'$. We use this as the slowest reaction constant in our BD simulations to probe the reaction-controlled limit. For weak adsorption ($\beta\eps \lesssim 3$), $\ktot_{\rm R}$ barely changes with $I$ as in the diffusion-controlled limit.  Nevertheless, the inhibition effect in the reaction-controlled scenario is also only very moderate for the strong adsorption cases ($\beta\eps \gtrsim 5$): The change of $\ktot_{\rm R}$ along with $I$ is maximally only up to two-fold at $\beta\eps = 10$. Moreover, reactions become faster with strong adsorption, which is opposite to the diffusion-controlled trends in \fig\ref{fig:kapp_k_inf}. In general, the agreement between simulation results and the theory is quite satisfying.

The weak variation of the rate with ionic strengths can be rationalized by inspecting the Langmuir isotherms, \Eq\ref{eq:surfcov}. For small adsorptions, the repulsive effect of the adsorbed charges is simply too small to affect the equilibrium distributions since the coverages are relatively small. \Figs S3(a) and S4(a) ($\beta\eps = 0.1$) show only step-like density profiles of reactant A around the nanoparticle without much accumulation for each $I$, implying minor inhibition effects.  For the stronger adsorption cases ($\beta\eps \gtrsim 5$), leading to higher coverages, the typical features of the Langmuir isotherm set in: the changes in the electrostatic contribution with varying ionic strength are the same in the numerator and denominator in the Langmuir equation (\Eq\ref{eq:surfcov}), and the trends somewhat cancel out. Hence, the inhibition effect is not that strong, as, e.g., in the diffusion-controlled scenario.  This can be also seen in the reactant and product coverages, (cf. for example \fig S6) where we find that the coverages do not change much with ionic strength. 

We note that we observe also a clear depletion zone in the density profiles of a width of the Debye screening length (e.g., ca. 1 nm for 100 mM salt concentration) away from the surface (see Fig.~S3 in the Supporting Information). This depletion zone reflects a kinetic barrier due to the electrostatic repulsion for the diffusive approach of the reactants. It will be important for more diffusion-influenced reactions (see below) but is not relevant in the reaction-controlled limit where the reactions are slow and the distributions always relax into equilibrium.

\subsection*{Diffusion-influenced regime}

We finally compare the simulation results with the interpolation formula for diffusion-influenced reactions, \Eq~\ref{eq:diffusion_influenced}.  The $\koff$ values needed to compute $\ktot_{\rm D}$ for the different interaction energies are taken from \fig\ref{fig:koff}. \fig\ref{fig:kapp_k_10kD} shows the simulation data for $\kR = 1.2~k_0'$ (open circles) and the diffusion-influenced theory (dashed lines). The simulation results are about 10-fold faster than the one in the reaction-controlled case ($\kR = 0.12~k_0'$, Fig.~\ref{fig:reaction_controlled}), but share qualitatively similar behavior: The product inhibition effect is still very modest for all adsorption energies $\eps$ and ionic strength $I$, and the total rate $\ktot$ increases with $\eps$. Importantly, we see that the empirical interpolation formula fits all data well. This becomes evident if we compare to the reaction-controlled theory only, plotted as thin dotted lines in \fig\ref{fig:kapp_k_10kD} for $\kR = 1.2~k_0'$. In particular, for the highest studied adsorptions, the reaction-controlled theory \Eqs\ref{eq:surfcov}--\ref{eq:elsurf} overestimates the simulation results by up to one order of magnitude.
(We have also compared \Eq~\ref{eq:diffusion_influenced} to the slowest rate  $\kR = 0.12~k_0'$ and found, as expected, hardly any difference to the full reaction-controlled limit, cf. \fig S7 in the Supporting Information).  We finally note that for further increasing surface rates $\kR$  we expect the simulation results to collapse at relatively high total rates for all adsorptions before inverting to the scenario in \fig\ref{fig:kapp_k_inf}, where rates decrease with adsorption strengths $\eps$. This is a consequence of the interpolation formula~\Eq~\ref{eq:diffusion_influenced}. 

\section*{Conclusion}

We have studied the role of electrostatic interactions in the reaction inhibition by reactants and products adsorbed on a spherical nanocatalyst in catalyzed-unimolecular reactions. We have derived approximate closed-form equations to describe the inhibition for a wide range of reaction speeds (diffusion-controlled to reaction-controlled), adsorption affinities, and salt concentrations. The equations are approximate but for these reasons sufficiently simple to be readily applied to fit and describe experimental rate measurements.  The validity of the equations is demonstrated by particle-resolved reaction-diffusion simulations of unimolecular transformation (A$\rightarrow$B) reactions catalyzed on the surface of a spherical nanoparticle.  The simulations and the theories reveal that electrostatic product inhibition is an essential mechanism in the diffusion-controlled limit and enhanced by strong adsorption and low ionic strength. Yet, the inhibition effect is less drastic in the reaction-controlled limit owing to the low reaction propensity $\kr$ and cancellations in the Langmuir isotherms. 

Our study can be extended in the future in various ways:  The theory could benefit from non-linear electrostatic approaches including the adsorption energy and steric interactions among adsorbates, as well as image charge effects~\cite{petersen2018}, or crowding effects.\cite{lee2020} Also, product dissociation (escape) rates from the catalyst seem to be influenced by electrostatic interactions and cooperativity, which deserves a more detailed investigation.  
One could also consider the inclusion of the intrinsic surface charge of the (functionalized) nanocatalyst itself.~\cite{roy} Moreover, in ``smart'' catalytic systems, nanoparticles are embedded in responsive hydrogels which shelter and control the catalysis\cite{Rafa:review, herves2012, lu2009, lu2013,roa2017}. There, the turnover rate is strongly influenced by the partitioning of ionic reactants in the hydrogel and on the catalysts,~\cite{kanduc2019} which eventually needs to be included in rate theory. Finally, in experiments, the reactions are typically bimolecular (A+B$\rightarrow$C).\cite{herves2012} We recently demonstrated that bimolecular reactions catalyzed on NPs exhibit strong coverage fluctuations, reactant correlations, and dynamic instabilities in the diffusion-controlled limit,\cite{YiChenLin2020} and approximate analytical approaches have emerged~\cite{roa2017}. Electrostatic interactions between all reactants and products with either asymmetric or opposite charge could significantly alter the kinetic rates and instabilities.

\section*{Acknowledgments}
The authors would like to thank Won Kyu Kim, Daniel Besold, and Matthias Ballauff for insightful discussions. This project has received funding from the European Research Council (ERC) under the European Union's Horizon 2020 research and innovation programme (grant agreement Nr. 646659). The authors acknowledge support by the state of Baden-W{\"u}rttemberg through bwHPC and the German Research Foundation (DFG) through grant no INST 39/963-1 FUGG (bwForCluster NEMO).

\footnotesize
\setlength{\bibsep}{0pt}

\bibliography{ref}

\clearpage

\makeatletter
\setlength\acs@tocentry@height{8.25cm}
\setlength\acs@tocentry@width{4.45cm}
\makeatother

\section*{TOC Graphic}
\centering
\includegraphics[width = 8.1cm]{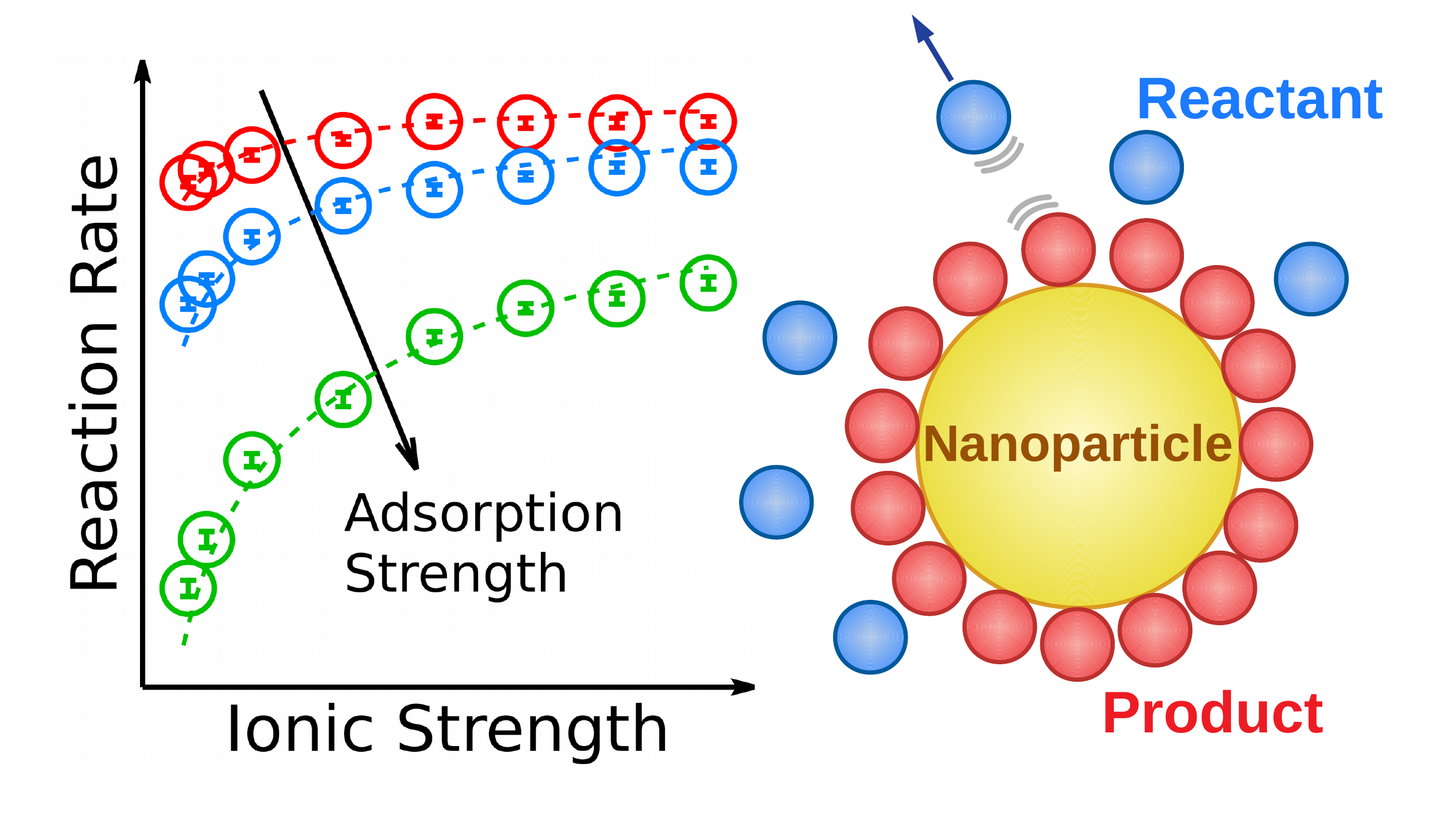}
\end{document}